\documentclass[showpacs,amsmath,amssymb,aps,showkeys,floatfix,prd,a4paper]{revtex4}
\usepackage[dvips]{graphicx}
\usepackage{dcolumn}
\usepackage{bm} 
\usepackage{epsfig}
\usepackage{amsfonts}
\usepackage{amssymb,amscd}
\usepackage{subfigure}
\usepackage{xcolor}
\usepackage{verbatim}
\newcommand{\NNN}{\mathcal{N}}
\newcommand{\kkk}{\mathcal{K}}
\newcommand{\rv}{\bm r}
\newcommand{\bv}{\bm b}
\newcommand{\dv}{\bm \Delta}

\newcommand{\Pelotas}{High and Medium Energy Group, Instituto de F\'{\i}sica e Matem\'atica,
             Universidade Federal de Pelotas\\
             Caixa Postal 354,  96010-900, Pelotas, RS, Brazil.}
\newcommand{\IFUSP}{Instituto de F\'{\i}sica, Universidade de S\~{a}o Paulo,
           C.P. 66318,  05315-970 S\~{a}o Paulo, SP, Brazil.}

\begin{document}

\title{Exclusive vector meson photoproduction with proton 
dissociation in photon-hadron interactions at the LHC}
\author{V.P. Gon\c{c}alves$^{1}$, F.S. Navarra$^{2}$ and D. Spiering$^{2}$}
\affiliation{$^{1}$ \Pelotas\\ $^{2}$ \IFUSP}

\begin{abstract}
At forward rapidities and high energies we expect to probe the non-linear 
regime of  Quantum Chromodynamics (QCD).
One of the most promising observables to constrain the QCD dynamics  
in this regime is exclusive vector meson photoproduction (EVMP). 
We study the EVMP in association with a leading baryon 
(product of the proton dissociation)
in photon-hadron interactions that take place in $pp$ and $p$Pb
collisions at large impact parameters. 
We present the rapidity distributions for $\rho$ and $J/\psi$           
photoproduction in association with a leading baryon (neutron and delta 
states) at LHC run 2 energies. 
Our results show that the $V + \Delta$ cross section is almost 30 \% of 
the $V + n$ one.
Our results also show that a future experimental analysis of these processes 
is, in principle, feasible and can be useful to study the leading particle 
production.
\end{abstract}


\maketitle

\section{Introduction}
\indent In collisions involving a proton, processes in which the proton 
dissociates
are very important. 
In $ep$ collisions they can significantly affect EVMP. 
In addition to the main reaction $e+p \to e+V+p$, 
a non-negligible fraction of vector mesons $V$ 
may come from the reaction with proton dissociation $e + p \to e + V + X$. 
In the latter reaction the proton dissociation would reduce the rapidity gap
expected in the former one. 
We are thus facing two important challenges: 
the experimental identification of EVMP 
and the quantitative estimate of the contribution of EVMP with the proton 
dissociation.\\

\indent We have recently studied \cite{ln_hh_upc}
one of the possible proton dissociation processes, 
where the proton dissociates into a leading neutron and a pion, 
with the former carrying a large fraction of the proton momentum.
In principle, the presence of a leading neutron in EVMP can be used to tag 
the event. In this work we will extend and complement our previous study
on EVMP and consider processes where the proton splits into $\Delta \pi$ states. 
These are, after $p \to n \pi$, the next most important proton dissociation 
process. More details can be found in Ref. \cite{lb_hh_upc}.

\section{Formalism}
\begin{figure}
  {\psfig{figure=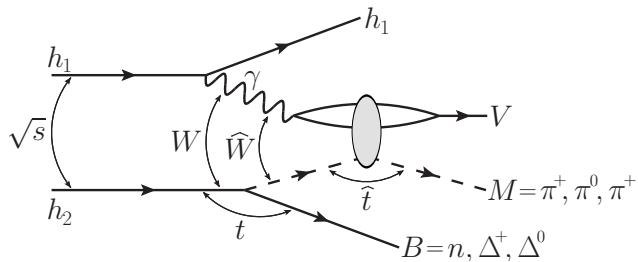,scale=0.45}}  
  \caption{Typical diagram of exclusive vector meson photoproduction in 
association with a leading baryon.}
  \label{diagram}
\end{figure}

The EVMP associated with a leading particle in photon-induced interactions in 
hadronic collisions is represented in Fig. \ref{diagram}.
This class of process can be factorized in terms of
the equivalent flux of photons ($N_{\gamma/h}$), 
the flux of virtual pions ($f_{\pi/h}$)
and the photon-pion cross section ($\sigma_{\gamma\pi}$) \cite{ln_hh_upc}:
\begin{eqnarray}
 \frac{d\sigma(h_1\!+\!h_2\!\rightarrow\!h_3\!+\!V\!+\!\pi\!+\!B)}{dYdx_Ldt} \label{sigma}
  &=& \kkk \cdot N_{\gamma/h_1}(\omega_L)\cdot f_{\pi/h_2}(x_L,t)\cdot \sigma_{\gamma\pi}(Y)\\
  &+& \kkk \cdot N_{\gamma/h_2}(\omega_R)\cdot f_{\pi/h_1}(x_L,t)\cdot \sigma_{\gamma\pi}(Y) \nonumber
\end{eqnarray}
where $Y=\ln(2\omega/M_V)$ is the rapidity of the vector meson with mass $M_V$,
$x_L$ is the leading particle longitudinal momentum and 
$t$ is square of the four-momentum of the exchaged pion. 
Moreover, $h_3$ corresponds to the initial hadron ($h_1$ or $h_2$)
which has emitted the photon and $\omega$ is the energy of the photons  
emitted by  the $h_1$ ($\omega_L\propto e^{-Y}$) and the $h_2$ 
($\omega_R\propto e^{+Y}$) hadrons. The constant factor $\kkk$ represents the 
absorptive corrections associated to soft rescatterings.
Using the color dipole picture, the process shown in  Fig. \ref{diagram} can 
be regarded as a sequence of four factorizable subprocesses:
($i$) a photon is emitted by one of the incident hadrons,
($ii$) this photon fluctuates into a quark-antiquark pair (the color dipole),
($iii$) the color dipole interacts diffractively with the pion (emitted by 
the other hadron), and ($iv$) the vector meson and the leading baryon are 
formed.\\
\indent Now let us discuss each term of expression \eqref{sigma}.
The $N_{\gamma/h}(\omega)$ describes the equivalent flux of photons (with 
energy $\omega$) of the hadron $h$.
The photon flux associated to the proton and nucleus can be described by the 
Drees-Zeppenfeld \cite{drees_zeppenfeld}
and relativist point-like charge models \cite{photon_flux}, respectively.
The photon flux of a nucleus is enhanced by a factor $Z^2$ in comparison to 
the proton one. Because of this, in proton-nucleus collisions we will only 
consider the processes in which the photon is emitted by the nucleus and  the 
pion is emitted by the proton. The function $f_{\pi/h}(x_L,t)$ represents the 
flux of virtual pions emitted by the proton $h$ and it is given by \cite{koepf96}
\begin{eqnarray}
  f_{\pi/p} (x_L,t) = \frac{g_{p \pi B}^2}{16 \pi^2}
   \frac{\mathcal{B}(t,m_p,m_B)}{(t-m_{\pi}^2)^2} \left(1-x_L\right)^{1-2t}  
  \exp \left[ 2b (t-m_{\pi}^2) \right]
  \label{B_flux}  
\end{eqnarray}
where $g_{p \pi B}$ is the proton-pion-baryon coupling constant,
$m_{\pi}$ is the pion mass 
and $b = 0.3$ GeV$^{-2}$ is related to the $p \pi B$ form factor \cite{f3}.
The term $\mathcal{B}$ depends on the type of the  produced baryon: 
\begin{eqnarray}
  \mathcal{B}(t,m_p,m_B) = 
    \left\{ 
     \begin{array}{ll}
      -t + \left(m_n-m_p\right)^2\,, & \hspace{0.2cm}\text{for $B\!=\!n$,}\\
      \displaystyle\frac{\left[\left(m_{\Delta}+m_p\right)^2-t\right]^2\left[
\left(m_{\Delta}-m_p\right)^2-t\right]}{12\,m_p^2\,m_{\Delta}^2}\,,
       & \hspace{0.2cm}\text{for $B\!=\!\Delta$,}
     \end{array}
    \right.
\end{eqnarray}
where $m_p$, $m_n$ and $m_{\Delta}$ are the respective masses of the proton, 
neutron and delta. In our analysis we will assume that 
$g_{p\pi^+n} = 19.025$, $g_{p\pi^+\Delta^0} = 11.676$ and 
$g_{p\pi^0\Delta^+} =  16.512$ \cite{babi1}. 
The function $\sigma_{\gamma\pi}$ describes the cross section 
of the diffractive interaction $\gamma + \pi \rightarrow V + \pi$ and 
in the dipole formalism it is given by \cite{ln_ep_exc}: 
\begin{eqnarray}
  \sigma_{\gamma \pi \rightarrow V\pi} = \frac{1}{16\pi}  \int_{-\infty}^0 
  \left| \int d\alpha \, d^2\rv \, d^2\bv\, e^{-i[\bv-(1-\alpha)\rv]\cdot \dv} 
   \, (\Psi^{V*}\Psi) \,2 {\cal{N}}_\pi(\hat{x},\rv,\bv)
 \right|^2 \, d\hat{t}\,\,,
  \label{sctotal_intt}
\end{eqnarray}
where $\hat{t}=-\Delta^2$ denotes the transverse momentum lost by the outgoing pion
and $\hat x=M_V^2/\hat W^2$ is the scaled Bjorken variable.
There are three center-of-mass energies present in this scattering:
$\hat W^2 = (1-x_L)W^2$ of the photon-pion system,
$W^2=M_Ve^Y\sqrt{s}$ of the photon-proton system 
and $\sqrt{s}$ of the hadron-hadron system.
The dipole variables are:
$\alpha$ $(1-\alpha)$ is the longitudinal momentum fraction of the quark (antiquark), 
$\bv$ is impact parameter and $\rv$ is the size of the dipole.
Finally, $(\Psi^{V*}\Psi)$ denotes the overlap between 
the real photon and exclusive final state wave functions, 
which we assume to be given by the Gauss-LC model \cite{kowalski_d74}
and  $\NNN^\pi(\hat{x},\rv,\bv)$ is the imaginary part of the  
dipole-pion forward scattering amplitude. 
This quantity is directly related to the QCD dynamics at high energies. 
As in Ref. \cite{ln_ep_inc}, we will assume that $\NNN^\pi$ 
can be expressed in terms of the dipole-proton scattering amplitude $\NNN^p$:
\begin{eqnarray}
  \NNN^\pi (\hat{x}, \rv, \bv) = R_q \cdot \NNN^p (\hat{x}, \rv, \bv) \,\,,
  \label{doister}
\end{eqnarray}
with $R_q$ being a constant. Moreover, we  will assume that 
$\NNN^p(\hat{x},\rv,\bv)$ is given by the bCGC model \cite{kowalski_d74,amir}.

\section{Results}
\indent Following our previous paper \cite{ln_hh_upc},
we will assume that absorptive corrections can be represented by 
a ${\cal{K}}$ factor.
The range of possible values was fixed in Ref. \cite{ln_ep_exc}
using HERA data \cite{h1_ln_exc} on $\sigma (\gamma p \rightarrow \rho \pi n)$, 
and it is given by $({\cal{K}}_{min}, {\cal{K}}_{med}, {\cal{K}}_{max}) = 
(0.152, 0.179, 0.205)$.
As a consequence, instead of a single curve for the rapidity distribution we 
will obtain a band.  As for the constant $R_q$, we shall,  
as in Refs. \cite{ln_ep_exc,ln_hh_upc},  assume that $R_q = 2/3$, 
as expected from the additive quark model. 
As implemented in the analysis of the H1 Collaboration \cite{h1_ln_exc},
we will assume that $p_T = \sqrt{|t|} <  0.2$ GeV.

\begin{figure}
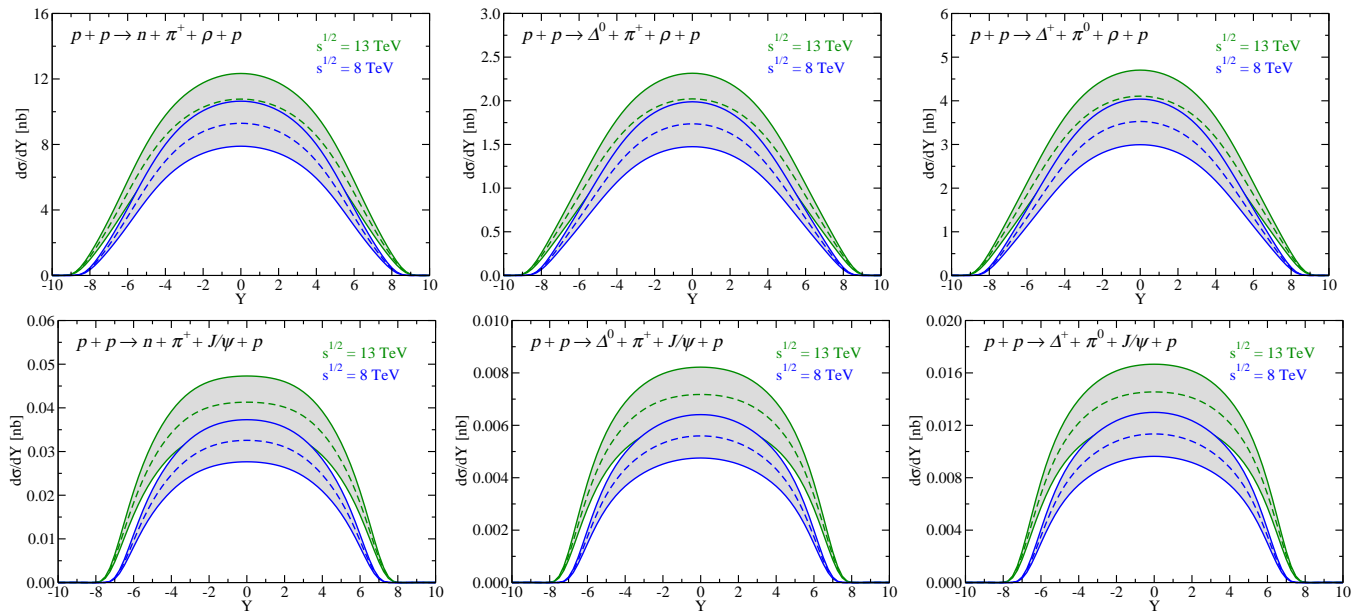

  \begin{tabular}{ccc}
    {\psfig{figure=LP_rho_pp_pion+leading_neutron_LHC,scale=0.23}} & 
    {\psfig{figure=LP_rho_pp_pion+leading_delta_zero_LHC,scale=0.23}} & 
    {\psfig{figure=LP_rho_pp_pion+leading_delta_plus_LHC,scale=0.23}} \\
    {\psfig{figure=LP_jpsi_pp_pion+leading_neutron_LHC,scale=0.23}} &
    {\psfig{figure=LP_jpsi_pp_pion+leading_delta_zero_LHC,scale=0.23}} &
    {\psfig{figure=LP_jpsi_pp_pion+leading_delta_plus_LHC,scale=0.23}}
  \end{tabular}
  \caption{Rapidity distribution for 
   $\rho$ and $J/\psi$
   production associated with a leading baryon $\gamma p$
   interactions at $pp$ collisions.}
  \label{pp}
\end{figure}

\indent Now we will present our predictions for the EVMP associated with a 
leading baryon
in photon-induced interactions considering $pp/p$Pb collisions at LHC energies.
In Fig. \ref{pp} we present our predictions for the rapidity distributions 
of the vector mesons $\rho$ and $J/\Psi$ produced in $pp$ collisions at 
$\sqrt{s} = 8 \text{ and } 13$ TeV. 
As expected from the symmetry of the initial state, 
the distributions are symmetric with respect to $Y=0$.
Moreover, the predictions for midrapidities $Y \approx 0$ increase with 
$\sqrt{s}$ and decrease with $M_V$. 
Additionally, the growth with the energy is faster for $J/\Psi$ than for 
$\rho$ production,
as expected from the partonic saturation physics \cite{kowalski_d74}.
Due to the fact that the $p\,\pi^0\Delta^+$ coupling constant is bigger than
the $p\,\pi^+\Delta^0$ one,
we can observe in Fig. \ref{pp} that the values of the cross sections
for leading $\Delta^+$ production are about two times bigger than
the values observed for leading $\Delta^0$ production. 

\begin{figure}
  \begin{tabular}{ccc}
    {\psfig{figure=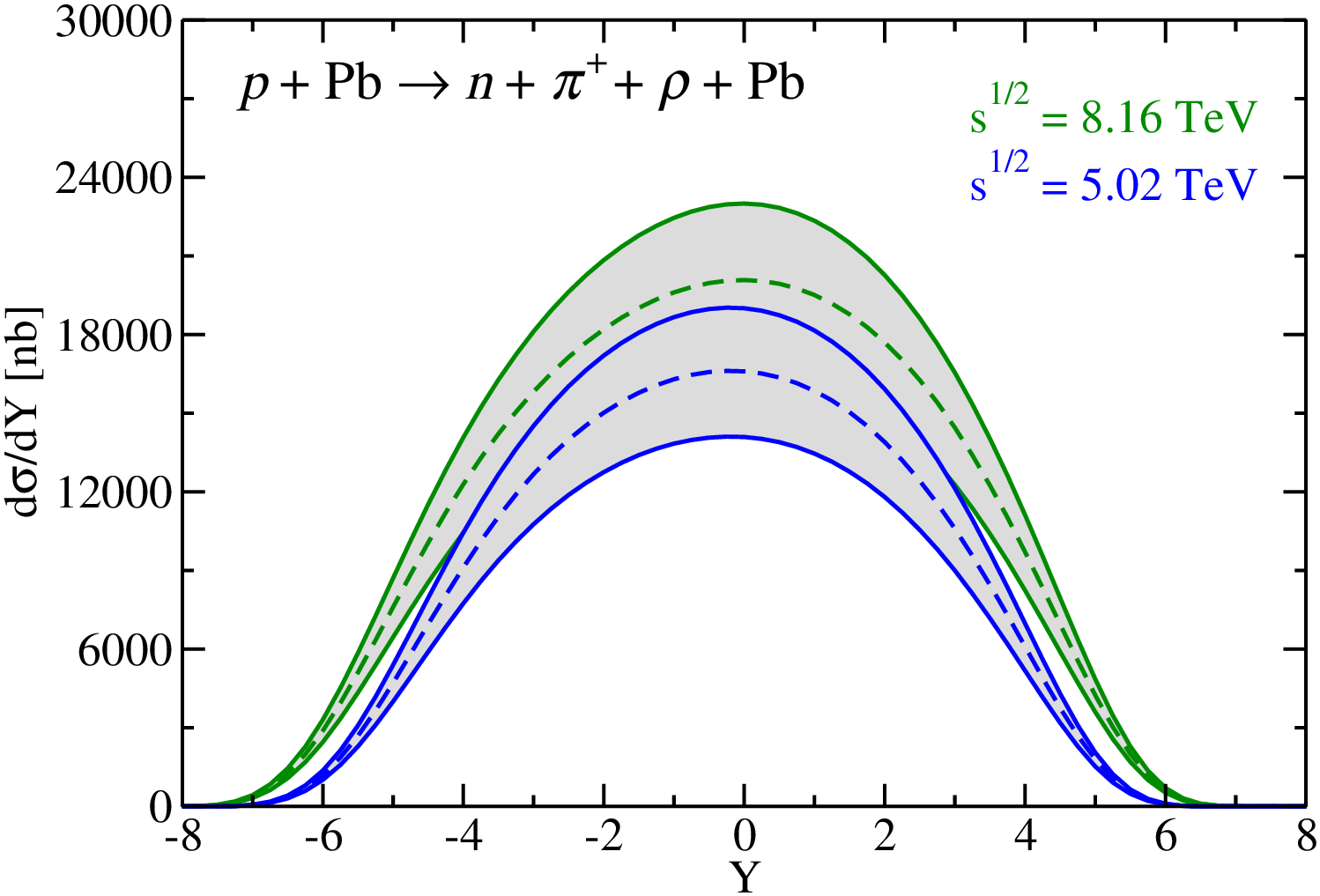,scale=0.23}} & 
    {\psfig{figure=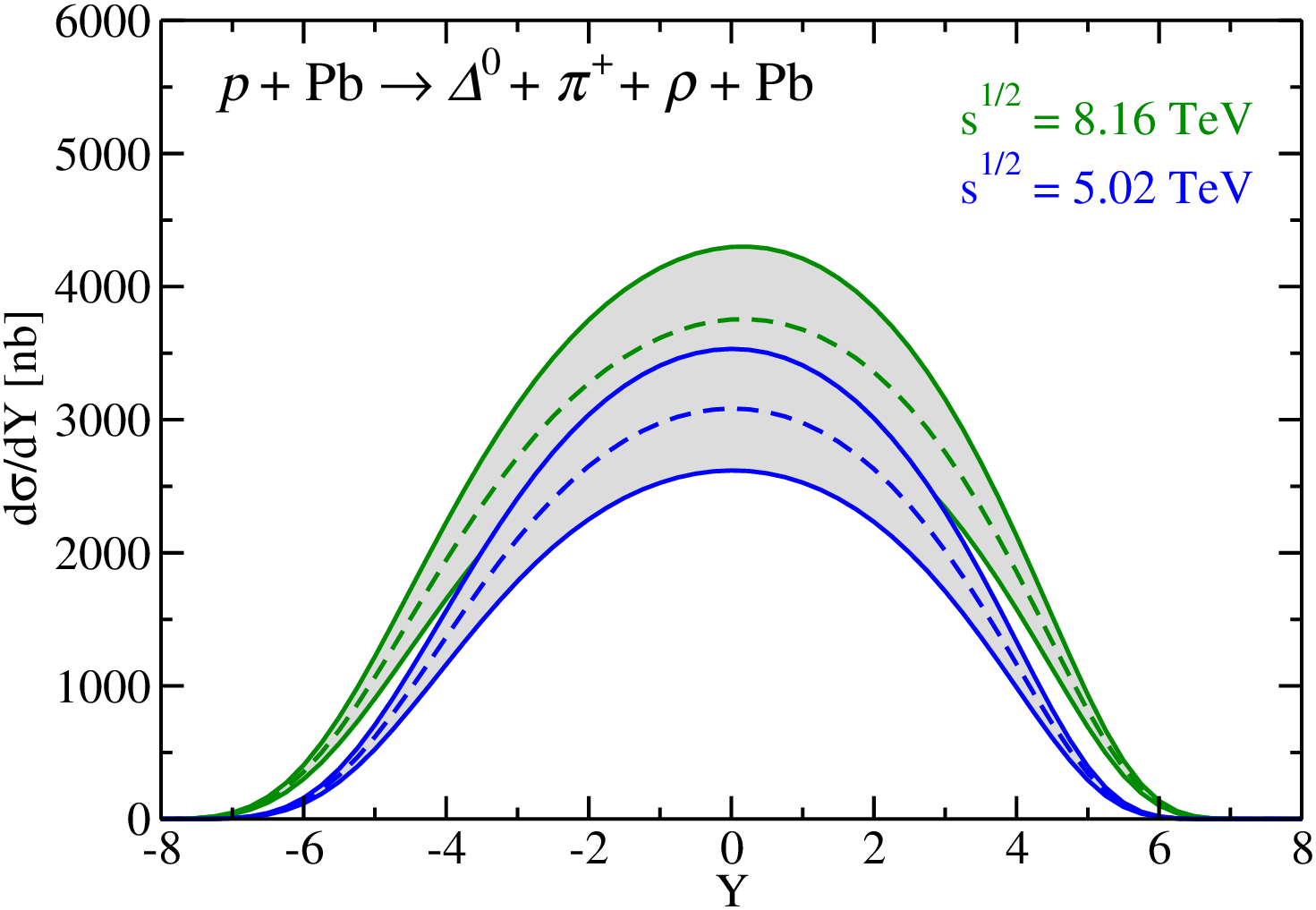,scale=0.23}} & 
    {\psfig{figure=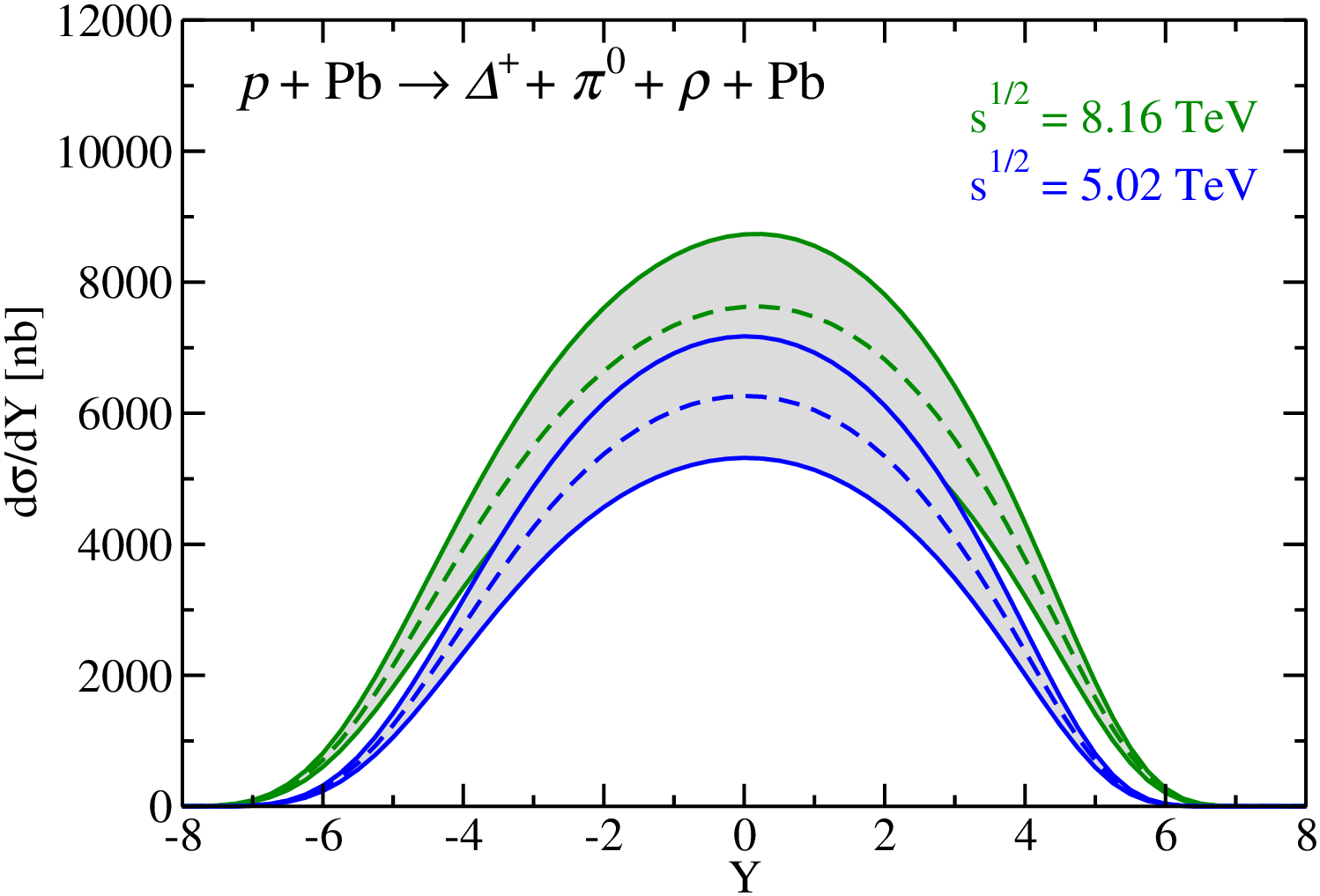,scale=0.23}} \\
    {\psfig{figure=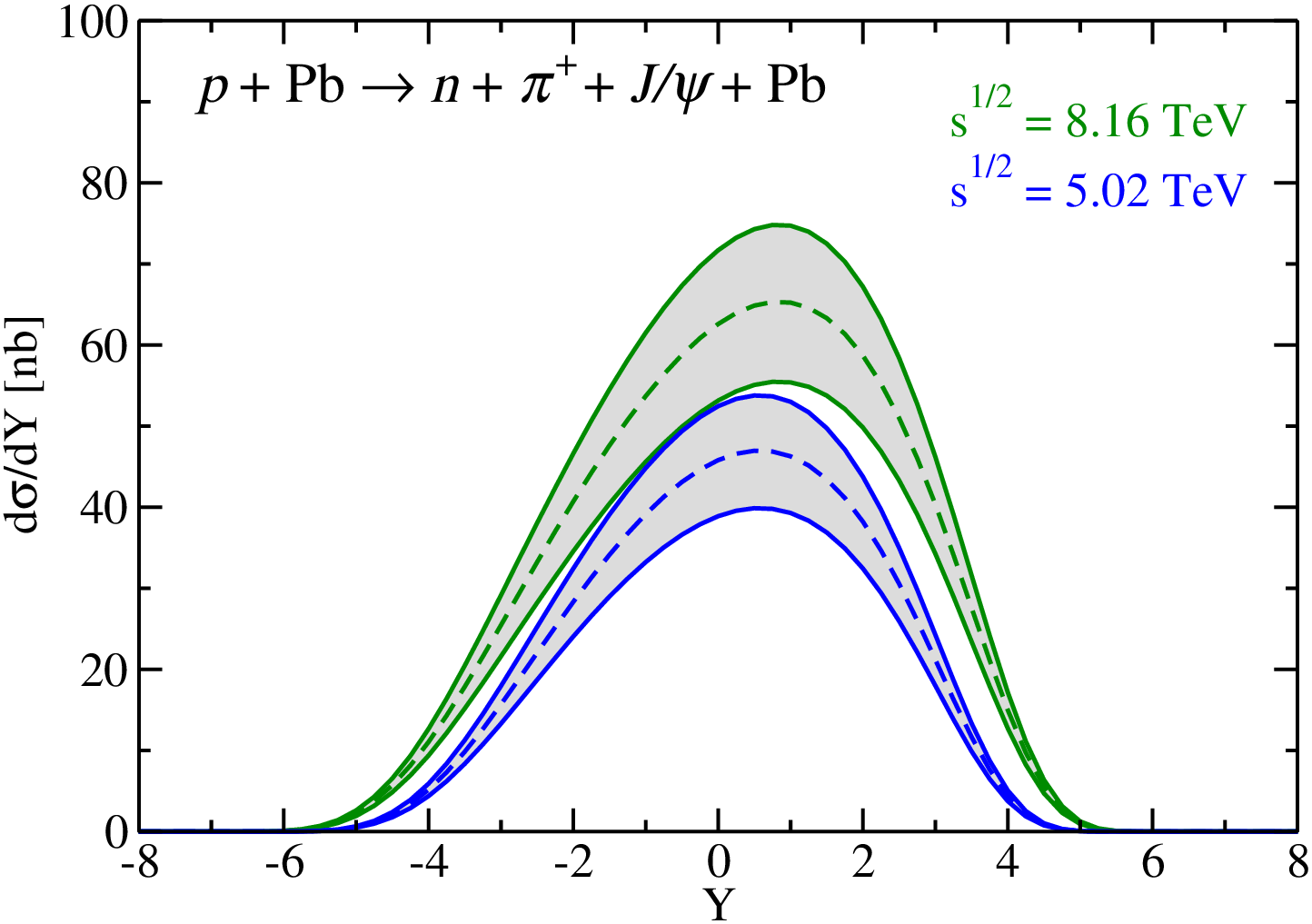,scale=0.23}} &
    {\psfig{figure=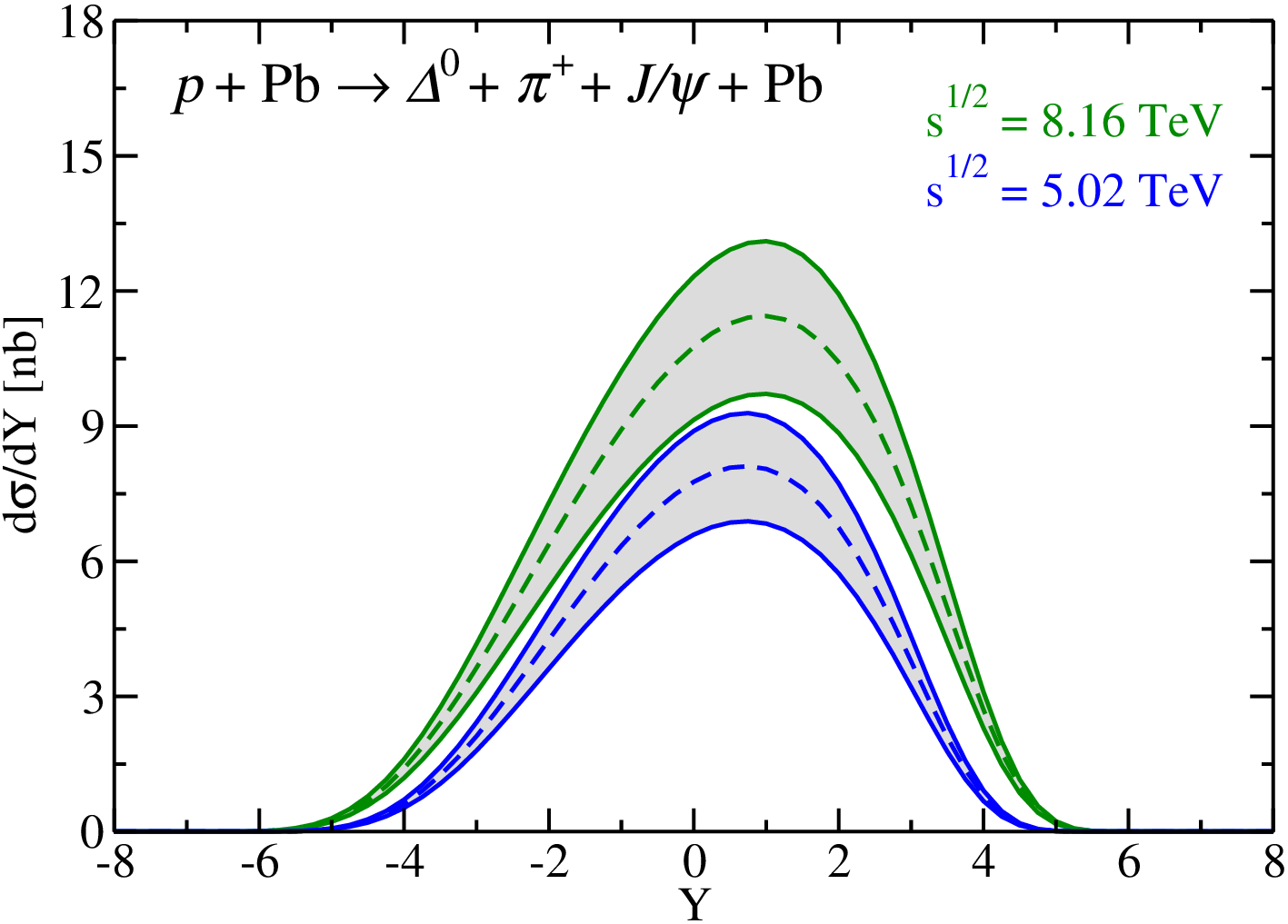,scale=0.23}} &
    {\psfig{figure=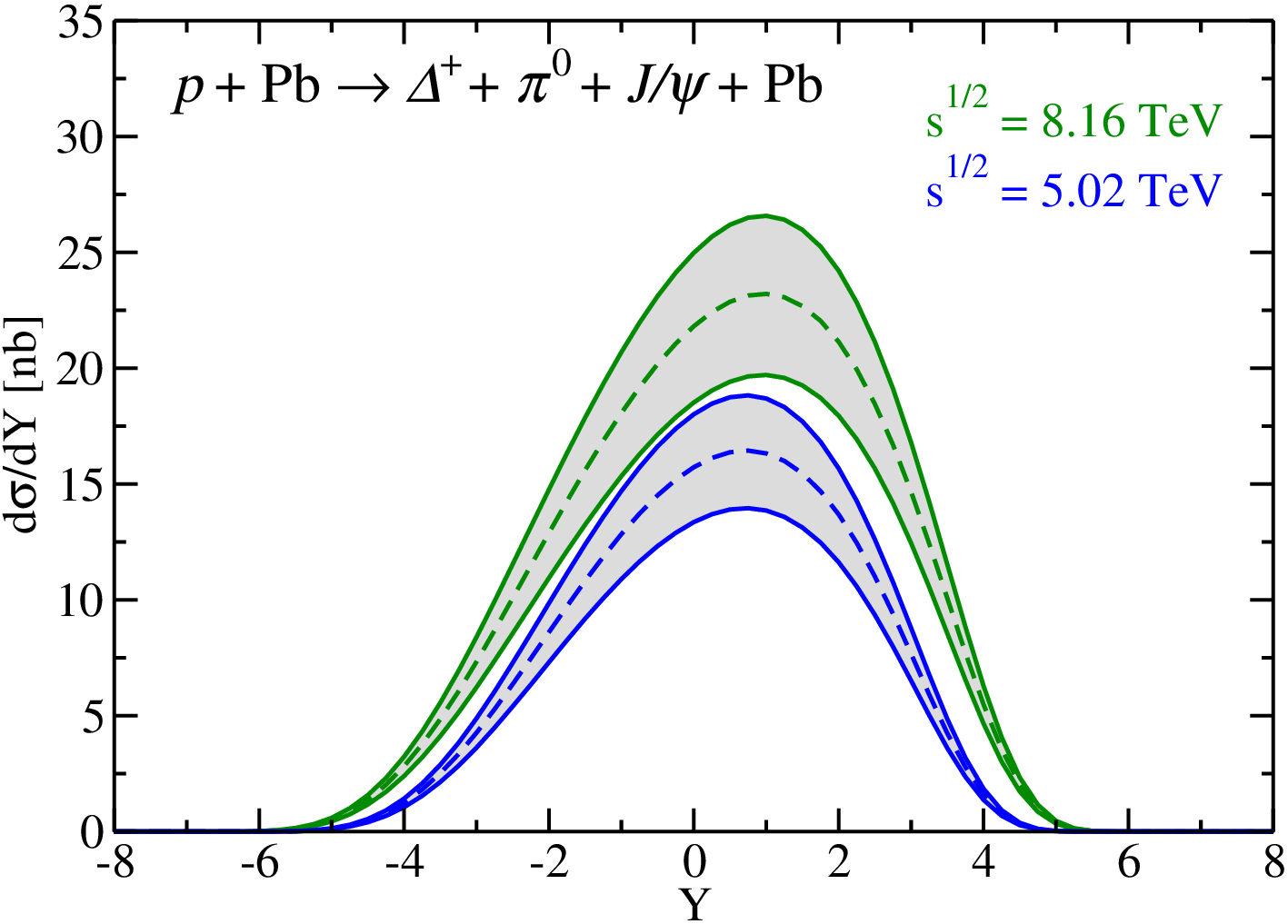,scale=0.23}}
  \end{tabular}
  \caption{Rapidity distribution for 
   $\rho$ and $J/\psi$
   production associated with a leading baryon $\gamma p$
   interactions at $p$Pb collisions.}
  \label{pA}
\end{figure}

\indent In Fig. \ref{pA} we present our predictions for the rapidity 
distributions 
of the vector mesons $\rho$ and $J/\Psi$ produced in $p$Pb collisions 
at $\sqrt{s} = 5.02 \text{ and } 8.16$ TeV.
In this case we have asymmetric distributions, 
since the photon-induced interactions are dominated by photons emitted by the nucleus.
Since $N_{\gamma/\text{Pb}} \gg N_{\gamma/p}$, 
the $p$Pb distributions are amplified in comparison with the $pp$ one.

\indent In the case of $\rho$ production associated with a leading neutron, 
we predict values of the order of $10^2$ ($10^5$) nb in $pp$ ($p$Pb) collisions.
Furthermore, the cross section of vector meson production associated with a  
leading delta is just one order of magnitude smaller, such that in the $\rho$ 
case we predict values of the order of $10^1$ ($10^4$) nb in $pp$ ($p$Pb) 
collisions at the same energies.
The decay channel $\Delta\rightarrow N\pi$  corresponds to 99.4\% of the 
delta decays \cite{pdg16} and hence  delta resonances can be an important 
background in processes with a leading nucleon.
In the case of measurements of vector meson production associated with a 
leading neutron,
the channels $\Delta^0\rightarrow n\,\pi^0$ and $\Delta^+\rightarrow n\,\pi^+$ 
together give a contribution of the order of $20\sim 28$\% of the direct 
leading neutron production. 
The secondary $\Delta$ decay into $n\pi$ will not affect the rapidity 
distribution of the vector meson,
but it will give a contribution to the neutron spectrum which is
softer than the one coming from the primary $p\rightarrow n\pi$ splitting.

\indent In comparison with the photoproduction of vector mesons (in $\gamma p$ 
interactions) in processes without the presence of a leading baryon \cite{evmp},
our predictions are smaller by approximately two (three) orders of magnitude 
for the leading neutron (delta), as expected from the experimental results 
obtained in $\gamma p$ collisions at HERA.   However, it is important to 
emphasize that these events will be characterized by  very forward baryons, 
which can be used to tag the events.

\section{Summary}
Recent results on photon-induced interactions at hadronic colliders have 
demonstrated that the analysis of these processes is feasible at the LHC.
This possibility has stimulated the improvement of the theoretical
description of these processes as well as the
proposal of complementary processes that also probe the QCD dynamics 
and are more easily tagged in collisions with a high pileup. 
Along this line, we have recently proposed the study of EVMP 
associated with a leading neutron in $\gamma p$ interactions at $pp$        
and $pA$  collisions and obtained large values for the total cross sections 
and event rates.
This result motivated the analysis performed in the present work, 
where we have extended the study to other leading particles,
with higher mass, which also generate a neutron in the final state through 
their decay. 
We have estimated the cross section for exclusive $\rho$ and $J/\psi$ 
photoproduction in association with a leading neutron and a leading 
$\Delta$ in $pp$ and $p$Pb collisions at LHC energies. 
We have found that the production associated with a leading $\Delta$ is 
non-negligible, being almost 30 \% of the one with a leading neutron. 
Our results indicate that the experimental analysis of this process is, in 
principle, feasible. 
In particular, if a combined analysis of the events using central
and forward detectors is performed, 
as those expected to occur using the CMS-TOTEM Precision Proton 
Spectrometer  and ATLAS + LHCf experiments. 
We expect thus that our results motivate a future experimental analysis 
of EVMP associated with a leading baryon in hadronic collisions at LHC.

\begin{acknowledgments}
This work was  partially financed by the Brazilian funding agencies 
CNPq, CAPES, FAPERGS and FAPESP. 
\end{acknowledgments}

\hspace{1.0cm}

\end{document}